\documentclass[
showpacs,floatfix,
twocolumn,%galley,
10pt,aip,apl,citeautoscript,superscriptaddress]{revtex4-1}

\usepackage{graphicx}
\usepackage{bm}
\usepackage{amsmath}
\usepackage{amssymb}

\begin{document}

\title{Generation of energy selective excitations in quantum Hall edge states}

\author{C. Leicht}
\affiliation{Physikalisch-Technische Bundesanstalt, 38116 Braunschweig, Germany.}

\author{P. Mirovsky}
\affiliation{Physikalisch-Technische Bundesanstalt, 38116 Braunschweig, Germany.}

\author{B. Kaestner}
\affiliation{Physikalisch-Technische Bundesanstalt, 38116 Braunschweig, Germany.}

\author{F. Hohls}
\affiliation{Physikalisch-Technische Bundesanstalt, 38116 Braunschweig, Germany.}

\author{V. Kashcheyevs}

\affiliation{Faculty of Physics and Mathematics, University of Latvia, Riga LV-1002 Latvia.}

\affiliation{Faculty of Computing, University of Latvia, Riga LV-1586, Latvia.}

\author{E. V. Kurganova}
\affiliation{High Field Magnet Laboratory, Institute for Molecules and Materials, 
Radboud University Nijmegen, 6525 ED Nijmegen, The Netherlands.}

\author{U. Zeitler}
\affiliation{High Field Magnet Laboratory, Institute for Molecules and Materials, 
Radboud University Nijmegen, 6525 ED Nijmegen, The Netherlands.}

\author{T. Weimann}
\affiliation{Physikalisch-Technische Bundesanstalt, 38116 Braunschweig, Germany.}

\author{K. Pierz}
\affiliation{Physikalisch-Technische Bundesanstalt, 38116 Braunschweig, Germany.}

\author{H. W. Schumacher}
\affiliation{Physikalisch-Technische Bundesanstalt, 38116 Braunschweig, Germany.}

%\date{\today}

\begin{abstract}

We operate an on-demand source of single electrons in high perpendicular
magnetic fields up to 30\,T, corresponding to a filling factor $\nu$ below
$1/3$. The device extracts and emits single charges at a tunable
energy from and to a two-dimensional electron gas, brought into well defined
integer and fractional
quantum Hall (QH) states. It can therefore be used for sensitive electrical
transport studies, e.g. of excitations and relaxation processes in QH edge
states.

\end{abstract}

\maketitle

%%%%%%%%%%%%%%%%%%%%%%%%%%%%%%%%%%%%%%%%%%%%%%%%%%%%%%%%
%% Motivation: sensitive and non-disturbing measurements
%%%%%%%%%%%%%%%%%%%%%%%%%%%%%%%%%%%%%%%%%%%%%%%%%%%%%%%%

Charge transport in two-dimensional electron gases placed in a strong
perpendicular magnetic field is ruled by chiral edge 
states~\cite{Halperin1982, Buttiker1988, Wen1990}. 
These edge states are now beeing exploited routinely in
fundamental physics experiments, e.g. in electron 
interferometers~\cite{Ji2003, Neder2007}. 
Moreover, gapless neutral edge excitations have been
predicted~\cite{Aleiner1994, Chamon1994}, though not yet directly observed in
experiments using quantum point contacts to generate edge excitations, as
performed in e.g. Refs.~\onlinecite{altimiras2010b} and~\onlinecite{Granger2009}. Additional
counterpropagating edge excitations in the fractional quantum Hall (QH) state
have also been predicted~\cite{Wen1990, MacDonald1990}, but were not found in
studies of edge magneto plasmons~\cite{Ashoori1992, Ernst1997}. Only very
recently a shot noise experiment found first indications for a neutral
counterpropagating mode~\cite{Bid2009}.

In this paper we demonstrate a new method to generate triggered single
energy selective excitations in integer and fractional QH edges to probe
possible edge excitations and relaxation processes. Furthermore,
this method allows to precisely control the emission statistics of the
electrons, which opens the possibility for efficient time resolved measurements.

\begin{figure}
  \includegraphics[width=7.0cm]{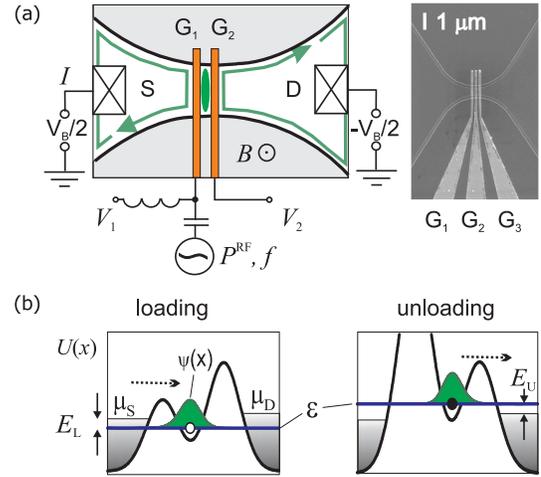}
  \caption{(color online) Description of the device and operating principle.
  (a) Left, schematic of the device.
  Electron micrograph of the sample shown on the right.
  (b) Schematic of the potential energy landscape along the channel for
  the stages of loading and unloading.}
  \label{fig:scheme}
\end{figure}

We adapt a structure that has 
previously been employed as high precision current
source, both in the dc~\cite{Kaestner2007c} and ac regime~\cite{feve2007}.
A schematic of our device and an electron micrograph are shown in Fig.~\ref{fig:scheme}a.
It was realized in an AlGaAs/GaAs heterostructure.
A 700$\,$nm wide constriction was wet-etched
inside a two-dimensional electron gas.
The device was contacted at source (S) and
drain (D) using an annealed layer of AuGeNi. The constriction
is crossed by Ti-Au finger gates G$_1$ and G$_2$. A quantum dot (QD) with a
quasibound state $\psi$ is formed by applying 
voltages $V_1$ and $V_2$ to G$_1$ and G$_2$, respectively;
a third gate G$_3$ is not used and set to ground.
The corresponding potential landscape along the constriction is shown in Fig.~\ref{fig:scheme}b.
An additional sinusoidal signal of power $P^{RF}$ and frequency $f$ is coupled to G$_1$
and varies both the height of the barrier and the energy 
$\varepsilon(t)= \varepsilon_1 \cos \omega t + \varepsilon_0$  
of the quasibound state ($\omega = 2 \pi f$).
During the first half cycle  $\varepsilon(t)$
drops below the chemical potential $\mu_S$ and $\psi$ is loaded with an
electron with energy $\mu_S -  E_L$ [see Fig.~\ref{fig:scheme}(b)].
During the second half-cycle, $\varepsilon (t)$ is raised sufficiently
fast above $\mu_D$ and the electron can be unloaded to the
drain with an excess energy $E_U$. This process, resulting into a \emph{quantized} 
current $I=e\cdot f$ with $e$ the electron charge,
is non-adiabatic and requires that the loaded QD state is raised sufficiently fast through the chemical
potentials $\mu_{S/D}$ to avoid unwanted charge transfer~\cite{Kaestner2007c}.
The scheme can be generalized to a quantized transport of $n$ electrons
per cycle, i.e. $I=n\cdot e\cdot f$, where $n$ can be derived from the 
decay cascade model~\cite{kaestner2010a}.

The current is accompanied by a periodic excitation in the drain at
energy $E_U$ above $\mu_D$. Upon application
of a strong perpendicular magnetic field $B$,
transport in S and D takes place
via edge channels, marked symbolically with green arrows in Fig.~\ref{fig:scheme}a.
Using the \emph{dynamical} QD it is now possible to trigger
single energy selective excitation quanta of the QH edge state.

\begin{figure}
  \includegraphics[width=8.0cm]{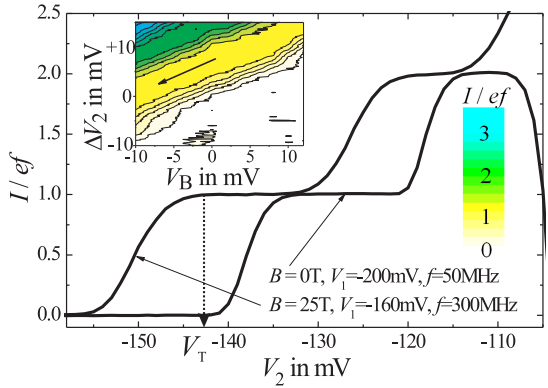}
  \caption{(color online) Normalized current $I/ef$ generated by electrons
  emitted into drain as function of $V_2$. The quantized regime $I = ef$ 
  under $B=0$ and $B=25\,$T conditions is obtained over a $V_2$ range of several mV.
  The threshold voltage $V_T$ is indicated by the dashed arrow, beyond which
  the quantized regime breaks down. Inset showing $I/ef$
  as the bias $V_B$ is varied.}
  \label{fig:tune}
\end{figure}

The number of electrons emitted into D per cycle may be tuned
using $V_2$, as shown in Fig.~\ref{fig:tune} for a measurement
carried out in a $^3$He cryostat with a base temperature of 300$\,$mK.
Under zero-field conditions approximately one electron is
emitted per cycle to D 
for $V_2 = -135 \cdots -120\,$mV. 
If a perpendicular magnetic field is applied it has been found
previously in acoustically driven dynamical QDs that quantization
is quenched for $B \geq 1\,$T~\cite{cunningham2}.
In the present case, where the dynamic
potential is generated directly by gates, quantization can be achieved up to
very high magnetic fields, as shown in Fig.~\ref{fig:tune}.
Here emission of single charges into D, i.e. quantized charge pumping, at $B=25\,$T is shown, corresponding to a fractional filling factor $\nu = 1/3$ of the undisturbed QH liquid.

The emission energy $ E_U$ depends on $f$ as well as on
the bias voltage, $V_B$, and $V_2$. It may be determined 
experimentally using gates
as energy filter, as for instance used in Ref.~\onlinecite{Hohls2006}. 
We have obtained a first estimate for $\Delta E =  E_U +  E_L$ 
at $B = 0$\,T based on the variation of the $e f$-plateau
lengths along $V_1$ as function of $P^{RF}$~\cite{kaestner2008}. 
For the studied device $\Delta E \approx 14\,$ to $17\,$meV
for frequencies ranging from 50$\,\cdots\,$300$\,$MHz, 
and for $V_2$ set to the negative side of the plateau 
($V_2 = V_T$, see Fig.~\ref{fig:tune}).

In the following we estimate the distribution of the energy 
of the emitted electrons, $p \left( E_U \right)$,
based on the Master equation model of Ref.~\onlinecite{Kaestner2007c}.
The sharpness of the distribution, $\Delta E_U$,
can be tuned by the \emph{selectivity} $s \equiv g / \varepsilon_1$ 
of the barrier at G$_2$ with 
$g \equiv \ln \Gamma_2^\mathrm{max} / \Gamma_2^\mathrm{min}$.
Here $\Gamma_2^\mathrm{max/min}$ are the
maximal and minimal tunnel rates during one cycle of modulation,
where we also assume that 
$\Gamma_2$ depends exponentially on $\varepsilon$.
To obtain an expression for the energy distribution we 
consider the case when unloading ($\varepsilon \ge \mu_D$) takes place during
the phase when $\varepsilon$ changes most rapidly,
such that $\varepsilon(t) \approx \varepsilon_1 \omega t + \mu_D$.
The problem can then be simplifed to a dynamic QD 
completely occupied at $t \rightarrow -\infty$ 
which unloads to drain via G$_2$ with increasing rate 
$\Gamma_2 (t) = \Gamma_2^0 \exp\left( \frac{1}{2}\omega\,g\,t \right)$,
where $\Gamma_2^0$ is the escape rate when $\varepsilon(t) = \mu_D$.
With these assumptions the distribution of the emission times is
peaked at $t_{e} = \beta^{-1}\ln\left( \beta/\Gamma_2^0 \right)$
with $\beta \equiv g\, \omega /2$. The width of the corresponding
energy distribution is then given by $\Delta E_U = 2\,/s$.
Hence, to obtain a narrow emission energy distribution
one may optimize the barrier shape of G$_2$ to maximize $s$.
The lowest achievable $\Delta E_U$ is limited by the 
quantum-mechanical uncertainty of energy, on the order of  
$\hbar \Gamma_2(t_e) = (g /2 ) \hbar \omega$.
For the frequencies chosen in this experiment 
the minimal $\Delta E_U$ lies in the $\mu e V$ range.

The derivation above also shows that the emission
energy $ E_U = \varepsilon(t_e) - \mu_D$
depends on the frequency $\omega$ 
and the tunnel rate $\Gamma_2^0$ logarithmically,
\begin{align} \label{eq:emienergy}
 E_U \approx \varepsilon_1 \omega t_e =
\Delta E_U\,\ln\left( \frac{\varepsilon_1\,\omega}{\Delta E_U\,\Gamma_2^0} \right) .
\end{align}
Since typically $\Gamma_2^0$ depends on $V_2$ exponentially,
the gate voltage can be readily used to tune the emission energy,
i.e. $E_U \propto - \,|e|\,V_2$.
To ensure single triggered excitation events (within a certain error margin) 
$V_2$ may be tuned only within the plateau voltage range, i.e. where
$I \approx e \, f$. 
The highest energy is obtained for the transition voltage, $V_T$, 
where $I = e f$ switches to $I = 0$, i.e. close to the negative side of 
the plateau where $\Gamma_2^0$ is minimal (see Fig.~\ref{fig:tune}).
From Eq.~\ref{eq:emienergy} it follows that
increasing the modulation amplitude $\varepsilon_1$ enhances 
$E_U$ only logarithmically. To extend the energy range
efficiently, the bias voltage 
$V_B \equiv \left( \mu_D - \mu_S \right)/ |e|$ may be made more negative, 
decreasing $\Gamma_2^0$ since the condition $\varepsilon =\mu_D$ 
will then take place earlier in the cycle, i.e. $E_U \propto -|e|\,V_B$.
At the same time the chance for emitting an additional
electron increases, as seen from the inset in 
Fig.~\ref{fig:tune}. This behaviour is consistent with the 
decay cascade model~\cite{kaestner2010a}, considering that
the time $t_c$ at which the decay cascade starts 
is given by $\varepsilon(t_c) \equiv \mu_S$. 
The corresponding escape rate at G$_1$, $\Gamma_1 (\varepsilon(t_c))$, 
controls the number of electrons captured per cycle.
To remain in the quantized regime, $V_2$ and consequently $\Gamma_2^0$ have to be 
decreased as indicated by the arrow in the inset of Fig.~\ref{fig:tune},
leading to an additional enhancement of $E_U$ according to Eq.~\ref{eq:emienergy}.
Hence, combining the $f$ -, $V_B$ - and $V_2$ - dependence an excitation energy
range up to several tens of meV should be possible using this technique.
Despite the potentially large energy, the heating of the edge state can be
kept at a minimum by choosing a sufficiently low frequency.
Furthermore, this energy selective and time controlled excitation source
could be combined with selective edge mode detection~\cite{Sukhodub2004}
and a time-gated detector technique~\cite{Ernst1997}
for sensitive studies of the underlying transport processes.

\begin{figure}
  \includegraphics[width=8.5cm]{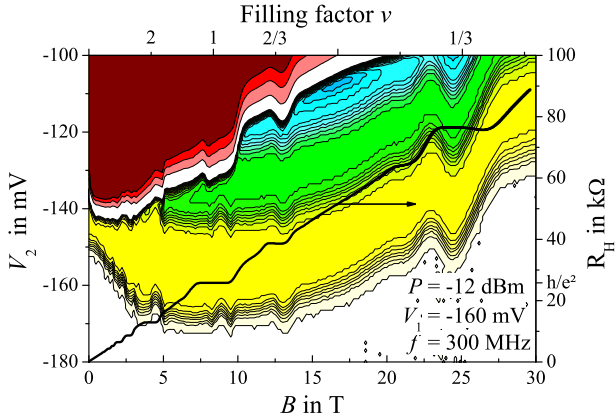}
  \caption{(color online) Normalized current $I/ef$ as a function of $V_2$ and $B$, as well as Hall resistance $R_H$ as a function of $B$. The power and frequency have been chosen to remain in the decay cascade regime~\cite{kaestner2010a}.
 The colors and contours in the diagramme correspond to those in Fig.~\ref{fig:tune}.}
  \label{fig:correlation}
\end{figure}

For the presented excitation source we require that the gates G$_1$ and G$_2$
coincide with the border of the undisturbed QH liquid, in order to
avoid broadening of the energy distribution $p(E_U)$.
In previous studies of this dynamical QD in perpendicular 
magentic field, such as in Refs.~\onlinecite{kaestner2009a} and~\onlinecite{wright2008},
the electron density of states and the corresponding filling factor
of the leads connecting to the QD via $G_1$ and $G_2$ could not be established.
In those works a wire of constant nominal width was employed, where
side wall depletion may result in varying electron densities 
inside the wire, different from the undisturbed QH liquid. 
The tapered channel geometry used in the present work
intends to avoid this complication.
Fig.~\ref{fig:correlation} shows evidence that in the case of the
tapered channel shape the QD extracts and emits electrons
directly from and to the undisturbed QH liquid. We conclude this from the
oscillations in $V_T(B)$ which coincide with the superimposed
Hall resistance $R_H$ determined for the undisturbed QH liquid. 
We relate these oscillations to the variations in $\mu_{S/D}$ 
for transitions between different integer and fractional filling factors~\cite{prange1986}
modifying the decay rates which $V_T$ is sensitive to.
In particular, the data in Fig.~\ref{fig:correlation} demonstrate the clocked
capturing of electrons directly from a fractional QH edge state.

Similar oscillations at lower magnetic fields have also been reported in Refs.~\onlinecite{cunningham2}
and~\onlinecite{wright2008}. In Ref.~\onlinecite{cunningham2} quantization was quenched
for $B \ge 1T$ and no clear comparison seems possible. The periodicity observed
in Ref.~\onlinecite{wright2008} does not correspond to the $R_H$ variation inferred from
the charge carrier density specified. This observation indicates emission into
a localized region of reduced electron density inside the etched channel.

Finally we note that charge pumping from fractional edge states as demonstrated here
may be developed further into the realization of a fractional charge pump as proposed by 
Simon~\cite{simon2000}, which may be used as a measurement of the charge of the fractional
quantum Hall quasiparticle.

We thank S. J. Wright, M. Kataoka, S. P. Giblin and J. D. Fletcher for
useful discussions. 
Technical support from H. Marx and U. Becker is gratefully acknowledged.
This work has been supported by 
%EURAMET joint research project with European Community's 7th Framework Programme, 
ERANET Plus under EU Grant Agreement No. 217257.
C.L. has been supported by International Graduate School of Metrology,
Braunschweig. Part of this work has been supported by EuroMagNET II under 
the EU contract number 228043.

%\bibliographystyle{apsrmp}

%\bibliography{literatureJabRef}

%merlin.mbs 2010-03-15 4.21a (PWD, AO, DPC)
%Control: key (0)
%Control: author (8) initials jnrlst
%Control: editor formatted (1) identically to author
%Control: production of article title (0) allowed
%Control: page (0) single
%Control: year (1) truncated
%Control: production of eprint (0) enabled
%

\end{document}